\documentclass[twocolumn,aps,prd,showpacs,superscriptaddress,floatfix]{revtex4}

\usepackage{bm}
\usepackage{dcolumn}
\usepackage{graphicx}
\usepackage{epsf}
\usepackage{epsfig}

\usepackage{amsmath}
\usepackage{amsfonts}
\usepackage{amssymb}
\usepackage{nicefrac}

\newcommand{\addrHD}{Max--Planck--Institut f\"ur Kernphysik,
Saupfercheckweg 1, 69117 Heidelberg, Germany}

\newcommand{\addrNIST}{National Institute of Standards and Technology,
Gaithersburg, Maryland 20899--8401}

\begin{document}

\bibliographystyle{myprsty}

\title{Self-energy values for $P$ states in hydrogen 
and low-$Z$ hydrogenlike ions}

\author{Ulrich~D.~Jentschura}
\affiliation{\addrHD}
\affiliation{\addrNIST}

\author{Peter~J.~Mohr}
\affiliation{\addrNIST}

\begin{abstract}
We describe a nonperturbative (in $Z\alpha$) 
numerical evaluation of the one-photon electron self
energy for $3P_{1/2}$, $3P_{3/2}$, $4P_{1/2}$ and 
$4P_{3/2}$ states in hydrogenlike atomic systems with 
charge numbers $Z=1$ to 5. The numerical results are 
found to be in agreement with known terms in 
the expansion of the self energy in
powers of $Z\alpha$ and lead to improved theoretical 
predictions for the self-energy shift of these states.
\end{abstract}


\maketitle

In this brief report, we consider the one-loop self-energy shift
for excited $P$ states in hydrogenlike systems.
High-accuracy numerical calculations of this 
effect are notoriously difficult especially in 
the region of low nuclear charge numbers $Z$,
because the renormalization entails a loss 
of up to nine decimal figures in the numerical 
calculation (the bound-electron self-energy is 
a residual effect which corresponds to the 
difference of the divergent self-energy 
of the bound electron, minus the corresponding, also 
divergent, shift for a free electron).

Furthermore, the calculations are needed for a 
self-consistent determination of fundamental constants,
which relies on the experimental/theoretical analysis
of a number of hydrogenic transitions.

One may obtain rather accurate values for the so-called self-energy 
remainder functions by an interpolation~\cite{Mo1975} of results 
for high and low nuclear charge numbers, which rely on two 
different approaches: (i) direct numerical calculations at high nuclear charge 
numbers (see e.g.~\cite{MoKi1992}, no $Z\alpha$-expansion), and
(ii) analytic calculations at low nuclear charge numbers,
employing the $Z\alpha$-expansion (see e.g.~\cite{JeEtAl2003}).
In this case, the values obtained for low nuclear charge
depend on the interpolation method used,
as well as (of course) on the reliability of both the 
numerical calculations at high $Z$ and the analytic calculations 
for low nuclear charge number. 

Here, we follow a third approach and calculate the self-energy 
without $Z\alpha$-expansion, at low $Z$, using a method described previously 
in~\cite{JeMoSo1999,JeMoSo2001pra,JeMo2004pra}. Essentially,
this method relies on an adequate formulation of the 
physical problem, by which divergent terms are suitably 
identified and calculated separately using semi-analytic 
approaches, and on the use of efficient numerical methods 
for the high-accuracy calculation of the Green function of the relativistic 
electron and for the 
evaluation of slowly convergent sums of intermediate 
angular momenta (it might be useful to mention the keyword ``convergence 
acceleration'' in that context~\cite{JeMoSoWe1999}).
On modern computer processors, it is not even necessary  
to parallelize the calculation; that latter approach had previously 
been employed in~\cite{JeMoSo1999,JeMoSo2001pra}.

All calculations reported here are carried out in the no-recoil limit,
which corresponds to an infinitely heavy nucleus. 
It might be instructive to recall the following 
subtlety, which is well known, and to include a slight detour:
At the current level of accuracy and especially at low $Z$, 
the reduced-mass dependence of the self energy
should be included for a comparison of the effect with experiment.
This can be done {\em a posteriori} by considering the semi-analytic expansion 
in Eq.~(\ref{defFLOnS}) below, using the formulas given 
in Eq.~(2.5b) of Ref.~\cite{SaYe1990}, which indicate the reduced-mass
dependence of the coefficients, and then the 
self-energy remainder values given in 
Tables~\ref{tableF3P12}---~\ref{tableF4P32} below in this 
brief report.

Returning to the discussion of our calculation,
we write the (real part of the) energy shift $\Delta E_{\rm SE}$ due to the
electron self-energy radiative correction as~\cite{SaYe1990}
\begin{equation}
\label{ESEasF}
\Delta E_{\rm SE} = \frac{\alpha}{\pi} \, 
\frac{(Z \alpha)^4 \, m_{\rm e} \, c^2}{n^3} \, 
F(nL_j,Z\alpha) \,,
\end{equation}
where $F$ is a dimensionless quantity ($m_{\rm e}$ is the 
electron mass, $\alpha$ is the fine-structure constant, 
and $c$ is the speed of light in vacuum).
In writing the expression $F(nL_j,Z\alpha)$, we follow the usual
spectroscopic notation for the quantum numbers of the single electron in 
a hydrogenlike ion; namely, we denote the principal quantum number
by $n$, the orbital angular momentum by $L$ and the
total electron angular momentum by $j$.  

\begin{table}[htb]
\begin{center}
\begin{minipage}{7.7cm}
\begin{center}
\caption{\label{tableF3P12} Numerical results for
the scaled self-energy function
$F$ ($3P_{1/2}, Z\alpha)$ and the self-energy remainder 
function $G_{\rm SE}$, as defined in Eq.~(\ref{ESEasF}),
in the regime of low nuclear charge numbers $Z$. See Fig.~\ref{fig1}.}
\begin{tabular}{l@{\hspace*{0.5cm}}r@{.}l@{\hspace*{0.5cm}}r@{.}l}
\hline
\hline
$Z$ &
\multicolumn{2}{l}{\rule[-3mm]{0mm}{8mm}
  $F(3P_{1/2},Z\alpha)$} &
\multicolumn{2}{l}{\rule[-3mm]{0mm}{8mm}
  $G_{\rm SE}(3P_{1/2},Z\alpha)$}\\
\hline
\rule[-1mm]{0mm}{6mm}
$1$ & $ -0$ & $115~459~16(5)$ & $ -1$ & $118~5(9)$ \\
\rule[-1mm]{0mm}{6mm}
$2$ & $ -0$ & $114~787~32(5)$ & $ -1$ & $089~7(2)$ \\
\rule[-1mm]{0mm}{6mm}
$3$ & $ -0$ & $113~831~60(5)$ & $ -1$ & $062~5(1)$ \\
\rule[-1mm]{0mm}{6mm}
$4$ & $ -0$ & $112~644~39(5)$ & $ -1$ & $036~30(6)$ \\
\rule[-3mm]{0mm}{8mm}
$5$ & $ -0$ & $111~258~78(5)$ & $ -1$ & $010~84(4)$ \\
\hline
\hline
\end{tabular}
\end{center}
\end{minipage}
\end{center}
\end{table}

\begin{table}[htb]
\begin{center}
\begin{minipage}{7.7cm}
\begin{center}
\caption{\label{tableF3P32} Same as Table~\ref{tableF3P12},
for the $3P_{3/2}$ state. See also Fig.~\ref{fig1}.}
\begin{tabular}{l@{\hspace*{0.5cm}}r@{.}l@{\hspace*{0.5cm}}r@{.}l}
\hline
\hline
$Z$ &
\multicolumn{2}{l}{\rule[-3mm]{0mm}{8mm}
  $F(3P_{3/2},Z\alpha)$} &
\multicolumn{2}{l}{\rule[-3mm]{0mm}{8mm}
  $G_{\rm SE}(3P_{3/2},Z\alpha)$}\\
\hline
\rule[-1mm]{0mm}{6mm}
$1$ & $  0$ & $134~414~38(5)$ & $ -0$ & $577~5(9)$ \\
\rule[-1mm]{0mm}{6mm}
$2$ & $  0$ & $134~792~58(5)$ & $ -0$ & $559~3(2)$ \\
\rule[-1mm]{0mm}{6mm}
$3$ & $  0$ & $135~332~45(5)$ & $ -0$ & $542~1(1)$ \\
\rule[-1mm]{0mm}{6mm}
$4$ & $  0$ & $136~006~06(5)$ & $ -0$ & $526~06(6)$ \\
\rule[-3mm]{0mm}{8mm}
$5$ & $  0$ & $136~795~05(5)$ & $ -0$ & $510~76(4)$ \\
\hline
\hline
\end{tabular}
\end{center}
\end{minipage}
\end{center}
\end{table}

\begin{table}[htb]
\begin{center}
\begin{minipage}{7.7cm}
\begin{center}
\caption{\label{tableF4P12} Same as Tables~\ref{tableF3P12} 
and~\ref{tableF3P32}, for the $4P_{1/2}$ state. See also Fig.~\ref{fig2}.}
\begin{tabular}{l@{\hspace*{0.5cm}}r@{.}l@{\hspace*{0.5cm}}r@{.}l}
\hline
\hline
$Z$ &
\multicolumn{2}{l}{\rule[-3mm]{0mm}{8mm}
  $F(4P_{1/2},Z\alpha)$} &
\multicolumn{2}{l}{\rule[-3mm]{0mm}{8mm}
  $G_{\rm SE}(4P_{1/2},Z\alpha)$}\\
\hline
\rule[-1mm]{0mm}{6mm}
$1$ & $ -0$ & $110~425~6(1)$ & $ -1$ & $164(2)$ \\
\rule[-1mm]{0mm}{6mm}
$2$ & $ -0$ & $109~720~3(1)$ & $ -1$ & $134~1(5)$ \\
\rule[-1mm]{0mm}{6mm}
$3$ & $ -0$ & $108~717~9(1)$ & $ -1$ & $105~5(2)$ \\
\rule[-1mm]{0mm}{6mm}
$4$ & $ -0$ & $107~471~8(1)$ & $ -1$ & $078~1(1)$ \\
\rule[-3mm]{0mm}{8mm}
$5$ & $ -0$ & $106~016~8(1)$ & $ -1$ & $051~20(8)$ \\
\hline
\hline
\end{tabular}
\end{center}
\end{minipage}
\end{center}
\end{table}

\begin{table}[htb]
\begin{center}
\begin{minipage}{7.7cm}
\begin{center}
\caption{\label{tableF4P32} Numerical results for
the scaled self-energy function
$F$ and the self-energy remainder
function $G_{\rm SE}$ for the $4P_{3/2}$ state. 
See also Fig.~\ref{fig2}.}
\begin{tabular}{l@{\hspace*{0.5cm}}r@{.}l@{\hspace*{0.5cm}}r@{.}l}
\hline
\hline
$Z$ &
\multicolumn{2}{l}{\rule[-3mm]{0mm}{8mm}
  $F(4P_{3/2},Z\alpha)$} &
\multicolumn{2}{l}{\rule[-3mm]{0mm}{8mm}
  $G_{\rm SE}(4P_{3/2},Z\alpha)$}\\
\hline
\rule[-1mm]{0mm}{6mm}
$1$ & $  0$ & $139~440~2(1)$ & $ -0$ & $609(2)$ \\
\rule[-1mm]{0mm}{6mm}
$2$ & $  0$ & $139~832~7(1)$ & $ -0$ & $590~4(5)$ \\
\rule[-1mm]{0mm}{6mm}
$3$ & $  0$ & $140~392~7(1)$ & $ -0$ & $572~9(2)$ \\
\rule[-1mm]{0mm}{6mm}
$4$ & $  0$ & $141~091~6(1)$ & $ -0$ & $555~5(1)$ \\
\rule[-3mm]{0mm}{8mm}
$5$ & $  0$ & $141~909~8(1)$ & $ -0$ & $539~41(8)$ \\
\hline
\hline
\end{tabular}
\end{center}
\end{minipage}
\end{center}
\end{table}

The leading terms 
in the semi-analytic expansion of $F(n{P}_j, Z\alpha)$
about $Z\alpha=0$ read
\begin{eqnarray}
\label{defFLOnS}
\lefteqn{F(n{P}_j, Z\alpha) = 
A_{40}(n{P}_j)  + \, (Z \alpha)^2} \, 
 \nonumber\\[2ex]
& & \times \left[A_{61}(n{P}_j) \,\ln(Z \alpha)^{-2} + 
G_{\rm SE}(n{P}_j, Z\alpha) \right]\,.
\end{eqnarray}
The $A$ coefficients have two indices, the first of which denotes the
power of $Z\alpha$ [including those powers explicitly shown in
Eq.~(\ref{ESEasF})], while the second index denotes the power of the
logarithm $\ln(Z \alpha)^{-2}$.  The evaluation of the coefficient 
\begin{equation}
\label{defa60}
A_{60}(nS_{1/2}) \equiv \lim_{Z\alpha \to 0} G_{\rm SE}(nS_{1/2},Z\alpha)
\end{equation}
has been historically problematic.

%
%
\begin{figure}[htb]
\begin{center}
%
\centerline{\mbox{\epsfysize=3.8cm\epsffile{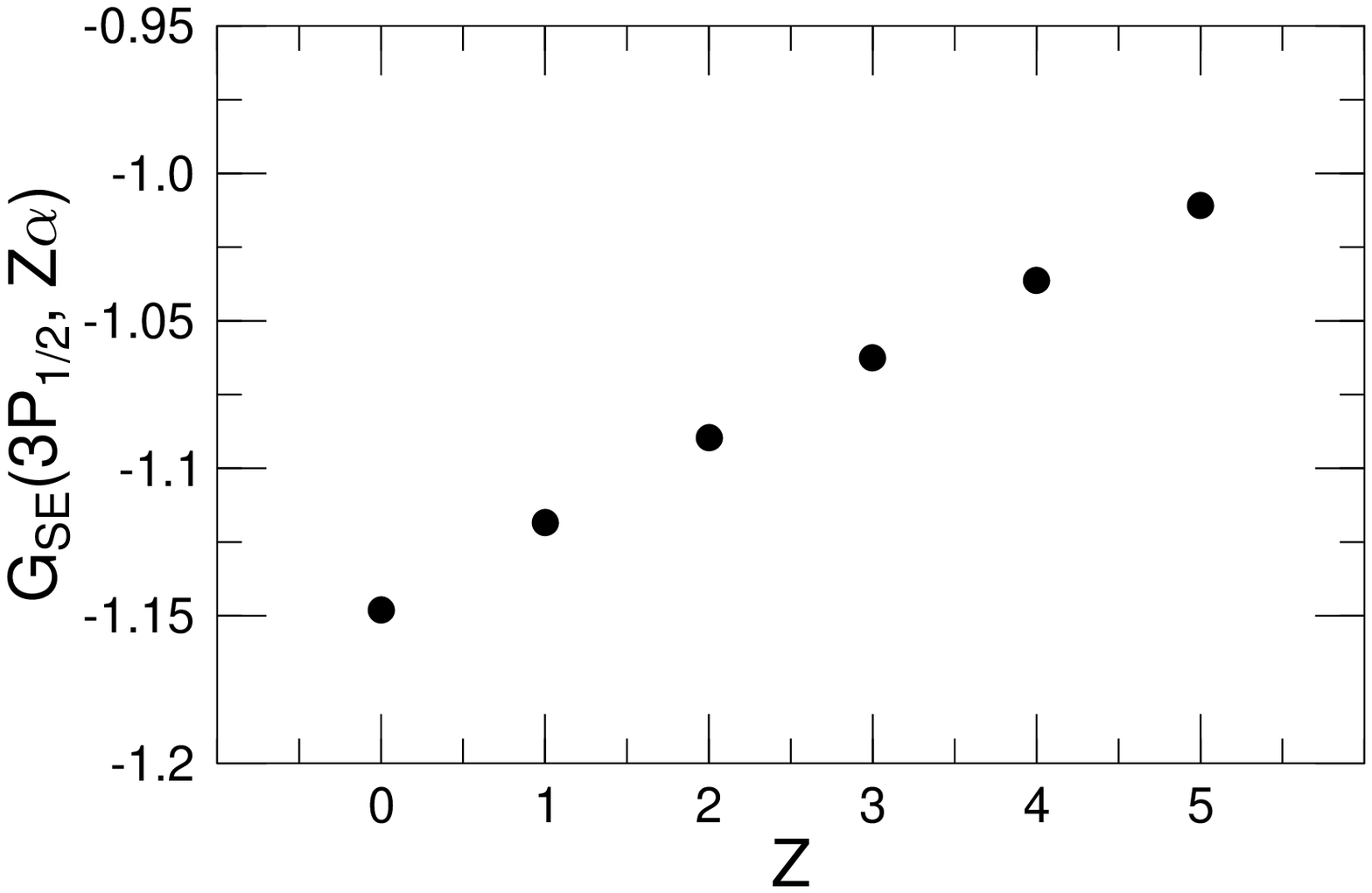}}}
\vspace{1cm}
\centerline{\mbox{\epsfysize=3.8cm\epsffile{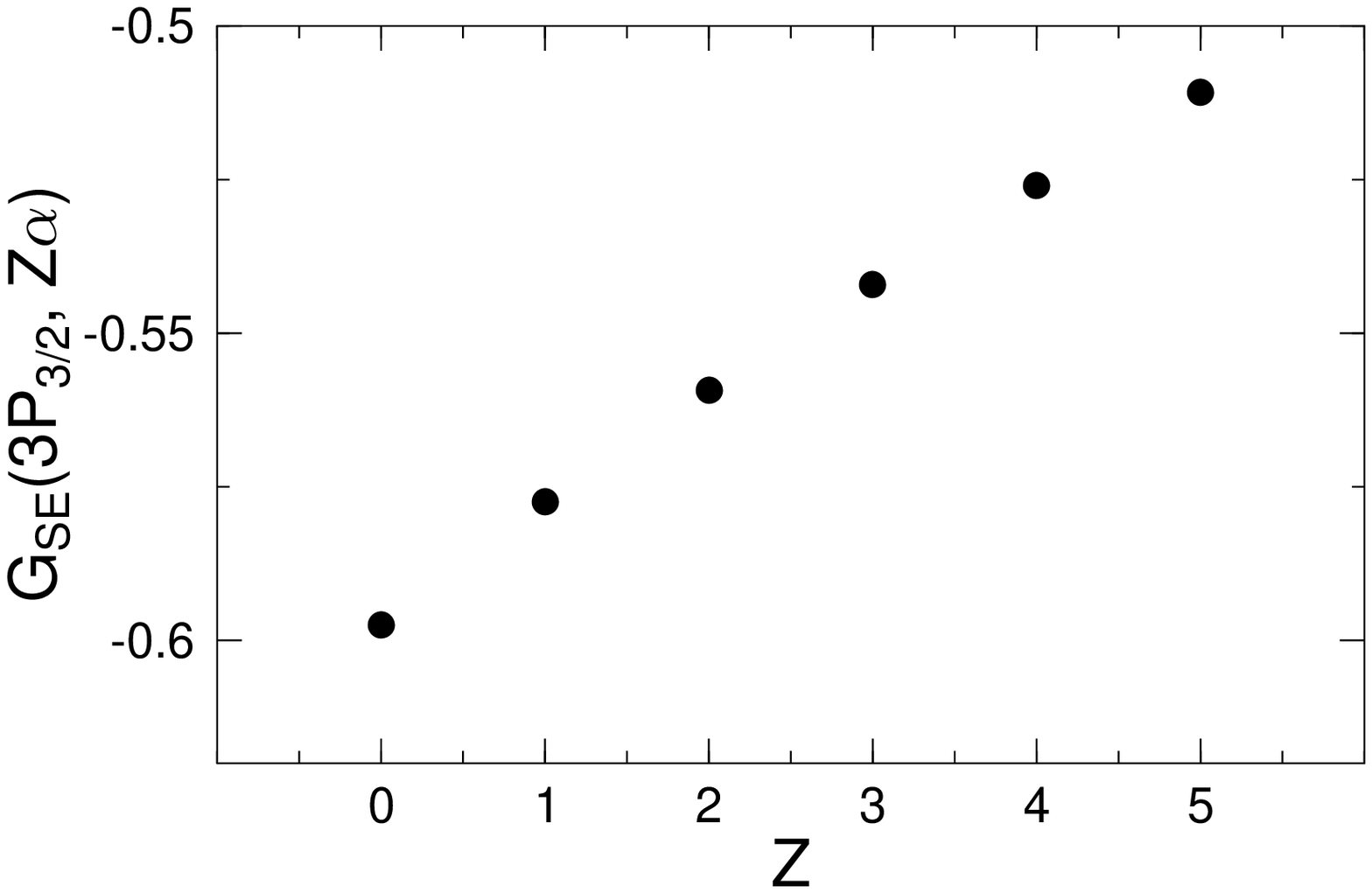}}}
\caption{\label{fig1} Comparison of the self-energy remainder
for $3P_{1/2}$ and $3P_{3/2}$ states, as listed in 
Tables~\ref{tableF3P12} and~\ref{tableF3P32}, to their low-$Z$ 
limit which is the $A_{60}$ coefficient [see Eqs.~(\ref{defa60})
and~(\ref{a60})]. Here, $Z$ is the nuclear charge, and 
the self-energy remainder $G_{\rm SE}$ is a dimensionless quantity.
The nuclear charge number is denoted by $Z$.}
\end{center}
\end{figure}

We now list the analytic coefficients and the
Bethe logarithms relevant to the atomic states under
investigation, referring the reader to~\cite{JeSoMo1997} 
for a more detailed discussion and further references,
\begin{subequations}
\begin{eqnarray}
\label{coeffs1S12}
A_{40}(n{P}_{1/2}) & = & -\frac16 - \frac43\,\ln k_0(nP)\,, \\[1ex]
A_{40}(n{P}_{3/2}) & = & \frac{1}{12} - \frac43\,\ln k_0(nP)\,.
\end{eqnarray}
\end{subequations}
Numerical values for the Bethe logarithms $\ln k_0(nP)$ are 
well known~\cite{KlMa1973,DrSw1990}.

The $A_{61}$-coefficients for the states under investigation read
\begin{subequations}
\begin{eqnarray}
A_{61}(3{P}_{1/2}) & = & \frac{268}{405} \,, \quad
A_{61}(3{P}_{3/2}) \; = \; \frac{148}{405} \,,\\[1ex]
A_{61}(4{P}_{1/2}) & = & \frac{499}{720} \,, \quad
A_{61}(4{P}_{3/2}) \; = \; \frac{137}{360} \,.
\end{eqnarray}
\end{subequations}
The $A_{60}$ coefficients have been evaluated in~\cite{JeSoMo1997},
and more recently in~\cite{JeEtAl2003} to an increased accuracy,
\begin{subequations}
\label{a60}
\begin{eqnarray}
A_{60}(3{P}_{1/2}) & = & -1.148\,189\,956(1)\,,\\[1ex]
A_{60}(3{P}_{3/2}) & = & -0.597\,569\,388(1)\,,\\[1ex]
A_{60}(4{P}_{1/2}) & = & -1.195\,688\,142(1)\,,\\[1ex]
A_{60}(4{P}_{3/2}) & = & -0.630\,945\,796(1) \,.
\end{eqnarray}
\end{subequations}
Note that the result for $3P_{1/2}$ had been given inaccurately
as $-1.147\,68(1)$ in~\cite{JeSoMo1997}.

%
%
\begin{figure}[htb]
\begin{center}
%
\centerline{\mbox{\epsfysize=3.8cm\epsffile{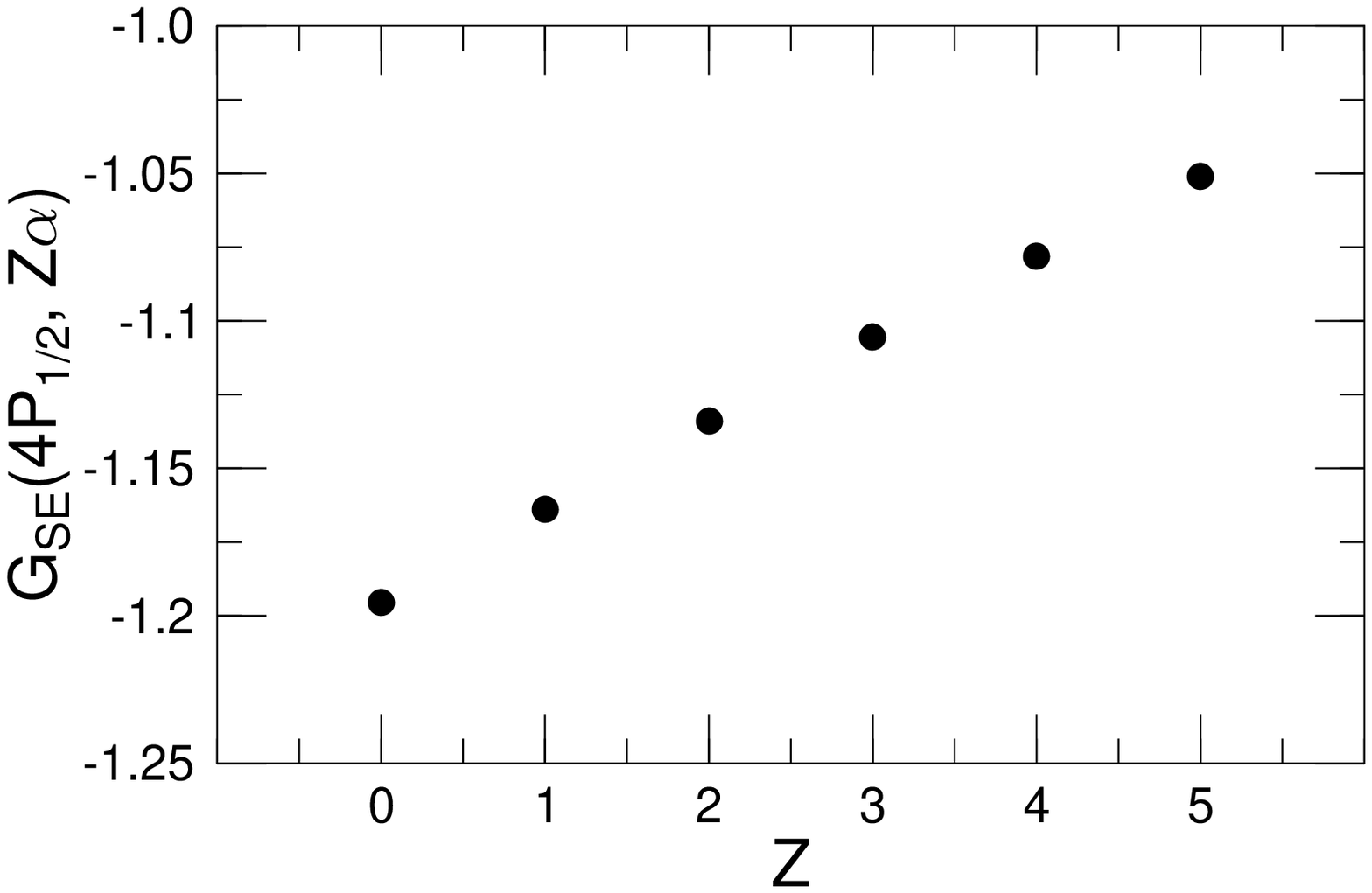}}}
\vspace{1cm}
\centerline{\mbox{\epsfysize=3.8cm\epsffile{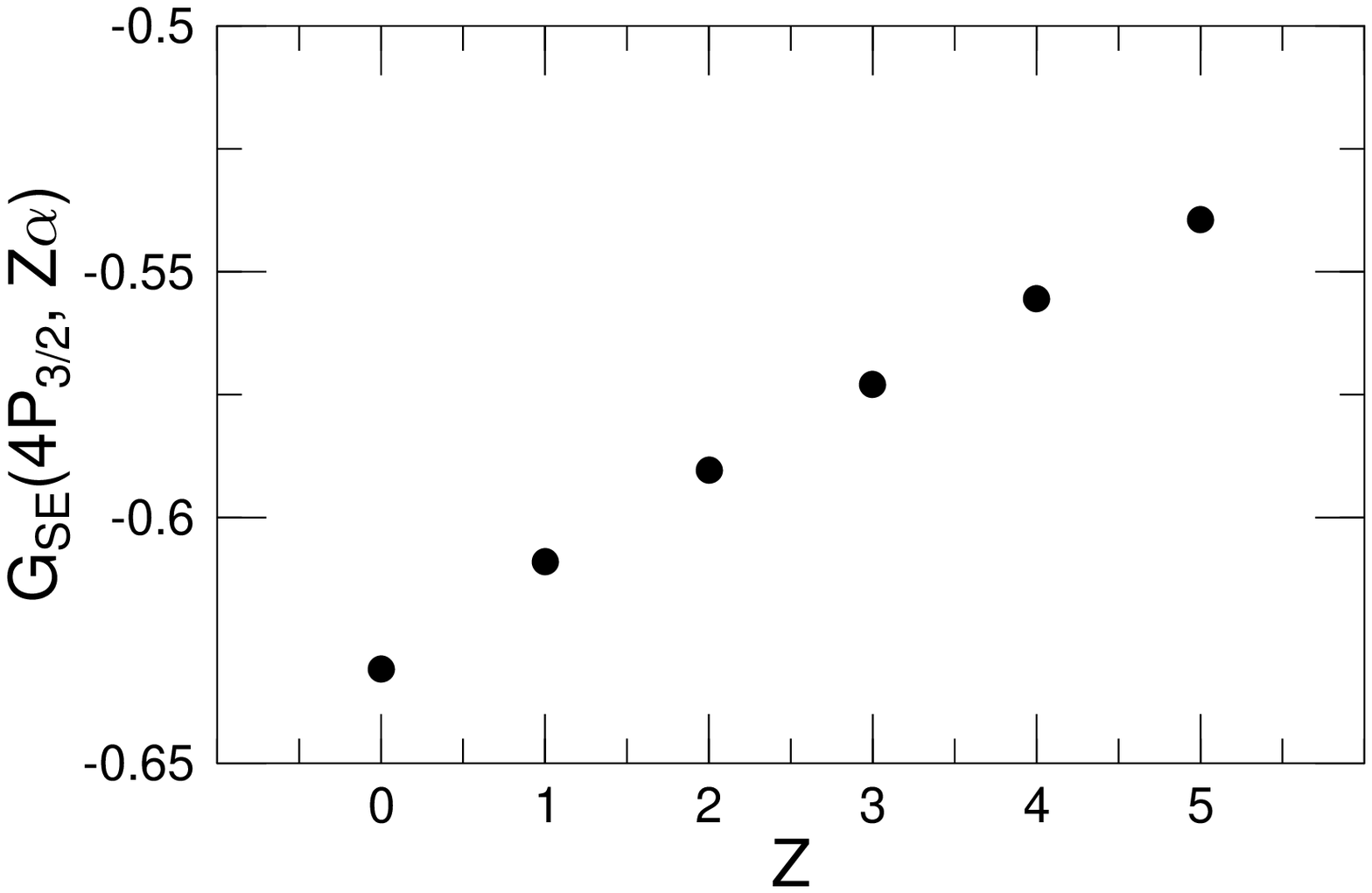}}}
\caption{\label{fig2} The analog of Fig.~\ref{fig1}
for $4P_{1/2}$ and $4P_{3/2}$ states.}
\end{center}
\end{figure}

Our calculation of the nonperturbative (in 
$Z\alpha$) electron self-energy for the $3{P}_j$ state 
(see Tables~\ref{tableF3P12} and~\ref{tableF3P32}) has 
a numerical uncertainty of 2~Hz in atomic
hydrogen. For the $4P_j$ states, the
numerical uncertainty is 1.3$\times Z^4$~Hz 
(see Table~\ref{tableF4P12} and~\ref{tableF4P32}).
In the non-recoil limit, our result for $4P_{1/2}$,
$Z=1$, corresponds to a self-energy shift of
\begin{equation}
\label{eq:F4P1_exact}
\Delta E_{\rm SE}(4P_{1/2}, Z=1) = -1\,404.239(1) \mbox{ kHz}\,,
\end{equation}
which is in agreement with the result 
$-1\,404.240(2) \mbox{kHz}$ obtained in~\cite{LBJeInMo2005}
via an interpolation of the low-$Z$ analytic 
results and high-$Z$ numerical data,
confirming (in this particular case) the validity 
of the interpolation procedure used for various 
excited hydrogenic states
in the latest adjustment of the fundamental physical 
constants~\cite{MoTa2005}.
Indeed, all entries for the self-energy remainder 
function $G_{\rm SE}$ in Tables~\ref{tableF3P12}---\ref{tableF4P32}
are in agreement with those used 
in~\cite{MoTa2005,JeKoLBMoTa2005} for the 
determination of the fundamental constants,
and for the precise calculation of hydrogenic energy levels using the
method of least squares. Our all-order
evaluation eliminates any uncertainty due to the unknown
higher-order analytic terms that contribute to 
the bound electron self-energy of $3P$ and $4P$ 
states [see Eq.~(\ref{defFLOnS})] and
improves our knowledge of the spectrum of
hydrogenlike atoms (e.g.~atomic hydrogen, He$^+$).
Furthermore, the numerical data for the self-energy remainders
{\em check} the validity of the highly involved analytic 
approach that led to the evaluation of the $A_{60}$-coefficients
as listed in Eq.~(\ref{a60}).

\section*{Acknowledgments}

U. D. J. thanks the National Institute of Standards and Technology for
kind hospitality during a number of extended research appointments.
The authors acknowledge E.-O.~LeBigot for help in obtaining numerical
results for selected partial contributions to the electron self-energy,
for the hydrogenic energy levels discussed in this work.

\end{document}